\documentclass[aps,preprint,nofootinbib]{revtex4}
\usepackage{amsfonts}
\usepackage{amsmath}
\usepackage{amssymb}
\usepackage{graphicx}%
\providecommand{\U}[1]{\protect\rule{.1in}{.1in}}

\begin{document}
\preprint{ }
\title[ ]{Internal conversion of nuclear transition in the
${}^{235m}U@C_{60-n}X_n$ molecules and related compounds}
\author{Alexei M. Frolov
\email{afrolov@uwo.ca}}
\affiliation{Department of Chemistry, University of Western Ontario, London K7L 3N6, Canada}
\date{\today}
\vspace{0.5cm}
\begin{abstract}
The internal conversion of nuclear transition in the ${}^{235}U$ atom is
considered. The low-energy $\gamma-$quanta ($E_{\gamma} \approx$ 76.8 $eV$)
are emitted during the $E3-$transition from the excited state ($I =
(\frac12)^+$) of the ${}^{235}U$ nucleus to its ground state ($I =
(\frac72)^-$). The decay rate of this ${}^{235m}U$ isomer ($E \approx 76.8$
$eV$) noticeably depends upon the chemical composition and actual physical
conditions (i.e. temperature $T$ and pressure $p$). By varying such a
composition and physical conditions one can change the life-time of the
${}^{235m}U$ isomer to relatively large/small values. A specific attention
is given to the fullerene molecules containing the central ${}^{235}U$
atom. It is shown that the decay rate $\lambda$ of the ${}^{235m}U$ isomer
in the ${}^{235m}U@C_{60-n}X_n$ molecules and related compounds can differ
significantly from the values obtained for isolated ${}^{235}U$ atoms. Some
applications of this effect are considered.
\end{abstract}

\maketitle

As is well known the rate of nuclear reactions and processes is usually
independent of the chemical background and physical conditions. In
particular, the transition rate between two arbitrary nuclear states cannot
depend upon the chemical environment and temperature (or pressure) of the
considered experimental sample. However, in some cases the nuclear
transition energies can be relatively small. If such an energy is comparable
to the energy of corresponding atomic levels, then the nuclear transition
can proceed by conversion of the emitted $\gamma-$quanta into electron
shells of the considered atom (Akhiezer and Beresteskii (1965) \cite{AB}).
Obviously, the most interesting case is the conversion of nuclear
transition to the outer electron shells of atoms. In these cases one can
change, in principle, the observed nuclear conversion rate by varying the
chemical environment and/or physical conditions. Such a situation can be
found in some heavy atoms, e.g., in the ${}^{235}U$ atom.

This effect was already experimentally demonstrated for the
${}^{235m}U$-isomer (M\'{e}vergnies (1969), (1972), Zhudov et al (1979)
\cite{Neve}, \cite{Nev2}, \cite{Zel}). The ${}^{235m}U$-isomer ($E \approx
76.8$ $eV$ $\pm 0.5$ $eV$ (Zhudov et al \cite{Zel})) is extensively produced
by $\alpha$-decay of ${}^{239}Pu$ in the core of nuclear warheads and
reactors. This isomer is the first excited state ($I = (\frac12)^+$) of the
${}^{235}U$ nucleus which is only $\approx 76.8$ $eV$ above its ground state
($I = (\frac72)^-$). The corresponding (nuclear) $E3-$transition to the
ground state of the ${}^{235}U$ nucleus (Grechukhin and Soldatov (1976)
\cite{GS1}) usually proceeds as an internal conversion of the nuclear
transition to the outer electron shells (i.e. to the $5f_{5/2}, 5f_{7/2},
6s_{1/2}, 6p_{1/2}, 6p_{3/2}, 6d_{3/2}, 6d_{5/2}, 7s_{1/2}$ shells) of the
${}^{235}U$ atom. The half-life of the ${}^{235m}U$ isomer is $\approx$ 26
$min$ (Zhudov et al (1979) \cite{Zel}). In earlier experiments the nuclei
of ${}^{235m}U$-isomer were implanted (M\'{e}vergnies (1969) \cite{Neve})
into various metallic foils. The considered cases (M\'{e}vergnies (1969)
\cite{Neve}) included the $Au, Pt, Cu, Ni, V$ foils. The decay rate
$\lambda$ of the ${}^{235m}U$ isomer was measured in each of these cases.
The maximal deviation between the results obtained with different metals was
found to be $\approx 5 \%$ (M\'{e}vergnies (1969), (1972) \cite{Neve},
\cite{Nev2}).

The approximate theory of internal conversion of nuclear $\gamma-$quanta
emitted during the $E3-$transition in the ${}^{235m}U$ nucleus was developed
in Grechukhin and Soldatov (1976) \cite{GS1}. The basic idea which
drastically simplifies calculation of the corresponding matrix element
(Akhiezer and Beresteskii (1965) \cite{AB}) is based on the fact that the
proton orbit inside of the nucleus has significantly shorter radius than the
corresponding electron orbits in the considered atom. In Grechukhin and
Soldatov (1976) \cite{GS1} the incident electron shells in the $U$ atom were
described with the use of Hartree-Fock-Slater relativistic approach. In this
approach a model with a spherically symmetric, central potential is used to
represent the electron-nucleus and direct electron-electron interaction. The
exchange electron-electron interaction is replaced by an approximate (or
effective) local, central potential. In the central potential approach the
incident (or bound) electron state is designated by the $n, j, \ell$
numbers, where $n$ is the principal quantum number, $\ell$ is the angular
momentum and $j$ represents the total momentum of the considered electron
shell. For a given value of $j$ we have $\ell = j \pm \frac12$. One of these
$\ell-$values is even and the other is odd. Therefore, the known values of
$j$ and parity of the considered state uniquely determine $\ell$. It was
shown in Grechukhin and Soldatov (1976) \cite{GS1} that for the
$E3-$transition of the ${}^{235}U$ nucleus from the isomer level with spin
$I_1$ to the ground level with spin $I_2$ the partial conversion
probability ($W$) per one electron in the $n j \ell$ state takes the form
\begin{eqnarray}
 W(E3;I_1 \rightarrow I_2; n_1,\ell_1,j_1; \epsilon,\ell_2,j_2) &=&
 \Bigl(\frac{2 I_2 + 1}{2 I_1 + 1} \Bigr) \mid \langle I_2 \mid\mid E3
 \mid\mid I_1 \rangle \mid^2 \cdot \Bigl(\frac{R_0}{a_0}\Bigr)^6 \cdot
 \frac{e^4 m_e}{\hbar^3} \cdot \\
 && w_e(E3;n_1,\ell_1,j_1;\epsilon_2,\ell_2,j_2) \; \; \; , \nonumber
\end{eqnarray}
where $R_0 \approx 1.26 A^{\frac13} \approx 7.775 \cdot 10^{-13}$ $cm$ is
the nuclear (${}^{235}U$) radius, $A = 235$ is the total number of nucleons
in the ${}^{235}U$ nucleus and $a_0 = 5.29177249 \cdot 10^{-9}$ $cm$ is the
Bohr radius. The Planck constant divided by $2 \pi$ is designated in Eq.(1)
by $\hbar$, $m_e$ is the electron mass and $e$ is the electron charge. In
fact, $\frac{e^4 m_e}{\hbar} = 4.13413733 \cdot 10^{16}$ $sec^{-1}$ is the
inverse atomic time (or basic atomic frequency in $Hz$). Also, in this
formula $w_e(E3;n_1,\ell_1,j_1;\epsilon_2,\ell_2,j_2)$ is the so-called
electronic factor which is determined from relativistic atomic calculations
(see below). In this study we shall assume that the central field
approximation can be used to represent both the incident and final atomic
states. The notation $\langle I_2 \mid\mid E3 \mid\mid I_1 \rangle$ in
Eq.(1) stands for the dimensionless reduced matrix element of the nuclear
$E3-$transition. This matrix element is of the form
\begin{eqnarray}
 \langle I_2 \mid\mid E3 \mid\mid I_1 \rangle =
 \frac{1}{C^{I_2M_2}_{I_1M_1;3M}} \langle I_2 M_2 \mid \sum^A_{i=1}
 q_i \Bigl( \frac{r_i}{R_0} \Bigr)^3 Y_{3 M}({\bf n}_i) \mid I_1 M_1 \rangle
\end{eqnarray}
where $C^{L_2M_2}_{L_1M_1;LM}$ is the corresponding Clebsch-Gordan
coefficient, $Y_{L M}({\bf n})$ are the spherical harmonics. Also, in this
equation $R_0$ is the radius of ${}^{235}U$ nucleus defined above, while
$r_i = \mid {\bf r}_i \mid$, where ${\bf r}_i$ is the nucleon radius-vector
($i = 1, \ldots, A$) in the nucleus. The $q_i$ is the charge of nucleon,
i.e., $q_i = 1$ for protons and $q_i = 0$ for neutrons. It is clear that
only protons contribute to the sum in Eq.(2). In fact, the numerical value
of the reduced nuclear matrix element $\langle I_2 \mid\mid E3 \mid\mid I_1
\rangle$ has been evaluated in earlier studies. In particular, in
Grechukhin and Soldatov (1976) \cite{GS1} it was shown that the numerical
value for this matrix element is bounded between $\approx$ 1.12 $\cdot
10^{-2}$ and $\approx$ 1.26 $\cdot 10^{-2}$. Below, we shall assume that
the $\langle I_2 \mid\mid E3 \mid\mid I_1 \rangle$ matrix element is 1.20
$\cdot 10^{-2}$.

Finally, the expression for the partial conversion probability takes the
form
\begin{eqnarray}
 W(E3;I_1 \rightarrow I_2; n_1,\ell_1,j_1;\epsilon_2,\ell_2,j_2) = 2.3955
 \cdot 10^{-10} \cdot w_e(E3;n_1,\ell_1,j_1;\epsilon_2,\ell_2,j_2) \; \; \;
 sec^{-1} \; \; \; .
\end{eqnarray}
Thus, the original nuclear-atomic problem is reduced to the computation of a
pure atomic matrix element $w_e(E3;n_1,\ell_1,j_1; \epsilon_2,\ell_2,j_2)$
which includes only atomic (or molecular) wave functions (see below). In
particular, for an isolated ${}^{235}U$ atom the matrix element
$w_e(E3;n_1,\ell_1,j_1;\epsilon_2,\ell_2,j_2)$ takes the form
\begin{eqnarray}
 && w_e(E L;n_1,\ell_1,j_1;\epsilon_2,\ell_2,j_2) = \frac{8 \pi}{p} \frac{2
 \ell_1 + 1}{(2 L + 1)^2} \sum_{j_2 \ell_2} (2 j_2 + 1) \mid
 C^{\ell_20}_{\ell_10;L0} R_h(L, \ell_2, j_1, \frac12; \ell_1, j_2) \mid^2
 \times \\
 && (1 + \gamma_2) \mid \int_0^{+\infty} r^{-(L+1)} dr
 [g_{n_1 \ell_1 j_1}(r) g_{\ell_2 j_2}(p r) + \sqrt{\frac{\gamma_2 -
 1}{\gamma_2 + 1}} f_{n_1 \ell_1 j_1}(r) f_{\ell_2 j_2}(p r)] \mid^2
 \nonumber \; \; \; ,
\end{eqnarray}
where $p = \mid {\bf p} \mid$ is the total momentum of the outgoing (free)
electron, while $\gamma_2 = \frac{\epsilon_2}{m c^2}$ is the Lorentz
$\gamma-$factor of this electron. Also, in this formula $R_h(L, \ell_2, j_1,
\frac12; \ell_1, j_2)$ is the corresponding Racah function which is simply
related to the Wigner 6-j symbol (see, e.g., Brink and Satchler \cite{BS}).
The radial functions $g_{n_1 \ell_1 j_1}(r)$ and $f_{n_1 \ell_1 j_1}(r)$ are
the large and small radial components of the bi-spinor wave function of the
bound $n_1, j_1, \ell_1$-state of the ${}^{235}U$ atom. Analogously, the
radial functions $g_{\ell_2 j_2}(p r)$ and $f_{\ell_2 j_2}(p r)$ are the
large and small radial components of the bi-spinor wave function of the
continuous atomic spectrum with energy $\epsilon_2$ and total momentum $p$
(Akhiezer and Beresteskii (1965) \cite{AB}). Note that $p$ is a scalar and
$\epsilon^2_2 = p^2 c^2 + m_e^2 c^4$. In fact, in the present case $L = 3$.
Therefore, in the case of $s-$ and $p-$ radial functions this radial
integral, in general, contains singularities.

The decay rate constant $\lambda({}^{235m}U)$ of the ${}^{235m}U$ isomer,
Eq.(3), is the sum of all partial conversion probabilities for
single-electron states. In the case of an isolated ${}^{235}U$ atom the
constant $\lambda({}^{235m}U)$ is written in the form
\begin{eqnarray}
 \lambda({}^{235m}U) &=& 2.3955 \cdot 10^{-10} \cdot \sum_{n_1,j_1,\ell_1}
 N(n_1,j_1,\ell_1) w_e(E L;n_1,\ell_1,j_1;\epsilon_2,\ell_2,j_2) \; \; \;
 sec^{-1} \; \; \; ,
\end{eqnarray}
where the sum is taken over all electron states (orbitals) in which the
conversion process is energetically allowed. Also, in this equation
$N(n_1,j_1,\ell_1)$ are the so-called occupation numbers of single-electron
atomic states $(n_1, j_1, \ell_1)$ (or orbitals). In general, by using a
pulse of laser radiation one can change the occupation numbers
$N(n_1,j_1,\ell_1)$ in the incident ${}^{235m}U$ atom. The decay rate
constant of the ${}^{235m}U$ isomer will change correspondingly. This method
can be used to measure the single-electron factors $w_e(E L;n_1,
\ell_1,j_1;\epsilon_2,\ell_2,j_2)$ experimentally. The theoretically
predicted values for all contributing single-electron shells are
$w_e(6p_{1/2}) = 4.95 \cdot 10^5, w_e(6p_{3/2}) = 2.17 \cdot 10^5,
w_e(6d_{5/2}) = 4.45 \cdot 10^4, w_e(6d_{3/2}) = 4.09 \cdot 10^4,
w_e(5f_{5/2}) = 6.87 \cdot 10^2, w_e(6s_{1/2}) = 6.58 \cdot 10^2,
w_e(5f_{7/2}) = 3.01 \cdot 10^2, w_e(7s_{1/2}) = 7.05 \cdot 10^1$. These
values have been obtained with the use of MOLFDIR package for relativistic
quantum chemistry calculations (Visscher et al (1994) \cite{MOLF}) and they
agree quite well with the corresponding $w_e$ factors determined in
Grechukhin and Soldatov (1976) \cite{GS1}. These values indicate clearly
that the electrons from $6p_{1/2}, 6p_{3/2}, 6d_{3/2}$ and $6d_{5/2}$ shells
of the ${}^{235}U$ atom are the main contributors to the ${}^{235m}U$ decay
rate constant. The decay rate constant $\lambda({}^{235m}U)$ computed with
our $w_e$ factors for the $(6s_{\frac12})^2 (6p_{\frac12})^2
(6p_{\frac32})^4 (5f_{\frac52})^3 (6d_{\frac32})^1 (7s_{\frac12})^2$
electron configuration of the ${}^{235m}U$ atom is $\approx$ 2193.18 $sec$
= 36.55 $min$. This value exceeds the known experimental value of the
$\lambda({}^{235m}U)$ constant (Zhudov et al (1979) \cite{Zel}) by
$\approx$ 40 \%.

Now, note that the same expression Eq.(5) can be used in those cases,
when the ${}^{235}U$ atom is bounded into a molecule or molecular
structure. In such cases, however, the numbers $N$ in Eq.(5) represent the
occupation numbers of molecular orbitals. In the first approximation
molecular orbitals can be considered as linear combinations of atomic
orbitals including single-electron orbitals form the central ${}^{235}U$
atom. Thus, in the case of a molecule which contains one ${}^{235}U$ atom
the last expression must be modified to the following form
\begin{eqnarray}
 && \lambda({}^{235m}U) = 2.3955 \cdot 10^{-10} \cdot \sum_{n_1,j_1,\ell_1}
 N^{(M)}(n_1,j_1,\ell_1) \sum_{\epsilon_2,\ell_2,j_2;F}
 w^{(M)}_e(n_1,\ell_1,j_1;A;\epsilon_2,\ell_2,j_2;F) \nonumber \\
 && = 2.3955 \cdot 10^{-10} \cdot \sum_{n_1,j_1,\ell_1}
 \Bigl[ N(n_1,j_1,\ell_1) + \Delta N^{(M)}(n_1,j_1,\ell_1) \Bigr]
 \sum_{\epsilon_2,\ell_2,j_2;F} \Bigl[ w_e(n_1,\ell_1,j_1;
 \epsilon_2,\ell_2,j_2) \\
 && + \Delta w^{(M)}_e(n_1,\ell_1,j_1;A; \epsilon_2,\ell_2,j_2;F) \Bigr]
 \; \; \; , \nonumber
\end{eqnarray}
where $(n_f, \ell_f, j_f)$ is the final state of the uranium atom. Here
$N^{(M)}(n_1,j_1,\ell_1)$ is the occupation number of the corresponding
atomic orbital in the ${}^{235}U$ atom which is bounded into a larger
molecular structure. For our present purposes it is important to note that
for the closed shells of the ${}^{235}U$ atom we always have
$N^{(M)}(n_1,j_1,\ell_1) < N(n_1,j_1,\ell_1)$, i.e., $\Delta N(n_1,j_1,
\ell_1) < 0$. This means that the occupation numbers for the bounded
uranium atom can be different from the occupation numbers of an isolated
${}^{235}U$ atom. In general, only by varying the occupation numbers of the
$6p_{1/2}$ and $6p_{3/2}$ orbitals in the ${}^{235}U$ atom one can change
the $\lambda({}^{235m}U)$ constant noticeably. The energy of the
$(6p)_{\frac12}-$electron in the uranium atom is $\approx$ 36.55 $eV$, while
the corresponding energies of $(6p)_{\frac32}-$electron is $\approx$ 26.80
$eV$. Therefore, a relatively large overlap between the $6p-$electrons of
uranium atom and surrounding molecular orbitals can be expected in those
cases when the considered molecule has a number of quasi-bound (or
resonance) excited states with close energies $\approx$ 26 - 37 $eV$. In
fact, it is shown below that the energies of excited molecular states can be
even $\approx$ 18 - 29 $eV$. However, it is clear that in any case such a
molecule must contain a relatively large number of atoms to avoid its
instant fragmentation.

The expression for single-electron conversion factor $w^{(M)}_e(E
L;n_1,\ell_1,j_1;A; \epsilon_2,\ell_2,j_2;F)$ in the case of a molecule with
one central ${}^{235}U$ atom takes a very complex form. For instance, the
indexes $A$ and $F$ which stand for the incident and final states of the
considered molecule are essentially the multi-indexes. This means that $A$
and $F$ contain all rotational, vibrational, electronic and spin quantum
numbers which are needed to represent uniformly the corresponding molecular
state. Below, we shall assume that the incident molecular state is an
excited molecular state and all molecular wave functions used below are
normalized to unity. As we mentioned above our present main interest is
related to the large molecules which contain one central uranium-235 atom.

An explicit expression for the $w^{(M)}_e(E L;n_1,\ell_1,j_1;A;
\epsilon_2,\ell_2,j_2;F)$ factor in Eq.(6) can be found, e.g., in the case
when the central ${}^{235}U$ atom is well separated from surrounding atoms
in the molecule. In this case the cluster approximation can be used and
expression for the additional $\Delta w^{(M)}_e(E L;n_1,\ell_1,j_1;A;
\epsilon_2,\ell_2,j_2;F)$ factor takes the from
\begin{eqnarray}
 \Delta w^{(M)}_e(n_1,\ell_1,j_1;A;\epsilon_2,\ell_2,j_2;F) = b \mid
 \sum^N_{j=1} \langle \Psi_A({\bf r}_1, \ldots {\bf r}_N) | B({\bf r}_j) |
 \Psi_F({\bf r}_1, \ldots {\bf r}_N) \rangle \mid^2 \; \; \; ,
\end{eqnarray}
where $b$ is a positive constant, ${\bf r}_j$ (where $j = 1, 2, \ldots, N)$
are the electron coordinates in the considered molecule. Here $N$ is the
total number of electrons in this molecule and $\Psi_A$ and $\Psi_F$ are the
incident and final molecular wave functions. In the first approximation the
operator $B({\bf r}_j)$ takes the following form
\begin{eqnarray}
 B({\bf r}_j) = \int_0^{\infty} g_{j_2 \ell_2}(p r) g_{n_1 j_1 \ell_1}(r)
 r^2 dr \oint \frac{1}{\mid {\bf
 r} - {\bf r}_j \mid} \Omega_{j_2 \ell_2 M_2}({\bf n}) \Omega_{j_1 \ell_1
 M_1}({\bf n}) d{\bf n} \; \; \; ,
\end{eqnarray}
where ${\bf r}$ is the electron radius for the considered single-electron
state (i.e. orbital) in the ${}^{235}U$ atom. The vector ${\bf n} =
\frac{{\bf r}}{r}$ is the corresponding unit vector, while $\Omega_{j_i
\ell_i M_i}({\bf n})$ ($i$ = 1, 2) are the `upper' spinors which depend upon
the angular variables ${\bf n}$. Also, in this equation $\gamma_2 =
\frac{\epsilon_2}{m c^2}$ and $p^2 = \frac{\epsilon^2_2}{c^2} - m_e^2 c^2$.
Note that the small radial components $f_{j_2 \ell_2}(p r), f_{n_1 j_1
\ell_1}(r)$ and corresponding angular spinors will contribute to the
$\Delta w^{(M)}_e(n_1,\ell_1,j_1;A;\epsilon_2,\ell_2,j_2;F)$ matrix element
only in the next (higher-order) approximation upon the fine structure
constant $\alpha \approx 7.29735308 \cdot 10^{-3} \ll 1$. In the higher
order approximation, however, the operator $B({\bf r}_j)$ takes
significantly more complicated form, since now it must include the
correction which correspond to the retarded interaction between charged
particles.

Note that the molecular conversion of nuclear transition represented by
Eq.(6) proceeds with the use of an intermediate atom (the ${}^{235}U$ atom
in our case). The direct molecular conversion of nuclear $\gamma-$quanta is
negligibly small, since the averaged molecular radius $R_M$ in large
molecules ($R_M \approx 10 a_0$) is significantly larger than the nuclear
radius $R_0$. This produces an additional factor of $\sim 10^{-6} - 10^{-8}$
in Eq.(1). In contrast with this, the internal molecular conversion of
nuclear transition with the use of intermediate ${}^{235}U$ atom has
significantly larger probability and can be observed experimentally.
Moreover, for some molecular structures the rate of molecular conversion of
nuclear transition can be different from the rate of pure atomic conversion
in the ${}^{235}U$ atom. This means a noticeable change in the decay rate
constant $\lambda({}^{235m}U)$ of the ${}^{235m}U$ isomer, Eq.(6), in some
molecules.

In general, the absorption of any significant amount of energy ($\approx
20 - 36$ $eV$ in the considered case) means, the partial (or complete)
dissociation of any few-atom molecule. In fact, below we shall consider
the molecular excitations with energies $\approx$ 77 $eV$ and even 100 $eV$.
The dissociation usually proceeds as a fragmentation of the incident
molecule into a number of fragments. Another possible way is the ionization
(or photoionization) of the incident molecule. In many cases, both molecular
fragmentation and photoionization occur together. In general, however, the
molecular bond strengths are $\approx 4.5$ $eV$, while the photoionization
of molecules requires $\approx 15$ $eV$. Moreover, the matrix element which
describes the photoionization contains the fine structure constant $\alpha
\approx 7.29735308 \cdot 10^{-3} \ll 1$. Therefore, the molecular
photoionization usually has smaller probability than the molecular
fragmentation. Below, our main interest is related to the consideration of
stable molecules in the incident and final state. In this case, one finds
that the minimal number $N_{min}$ of atoms in such a molecule must exceed
$N_{min} \approx 77 / 4.0 \approx 19$. In reality, the molecular
fragmentation starts at smaller energies ($\approx 2$ $eV$ per atom, see
below). This means that $N_{min} \approx$ 38 - 40, i.e. the molecules which
are of interest for our present purposes must contain at least 40 - 45
atoms. In the case of $\approx$ 100 $eV$ molecular excitation the number of
atoms per molecule must be $\approx$ 50 - 55.

To change the rate of internal conversion of low-energy nuclear
$\gamma-$quanta we propose to use the molecular structures based on
fullerenes $C_{60}, C_{84}, C_{100}$, etc. The corresponding molecular
structures have the following general formula ${}^{235m}U@C_n$, where $n
\geq 60$. Note that the fullerenes recently attracted a significant
experimental attention (see, e.g., Handschuh et al (1995), Joachim et al
(2000) \cite{Hand}, \cite{Joah} and references therein). In particular, the
fullerenes were suggested for numerous applications in the field of
molecular electronics (Park et al \cite{Park}). It is shown below that
fullerenes are also of certain interest for applied nuclear physics.

For our present purposes it is important to note that the instant
fragmentation of fullerenes and related molecular structures starts
(see, e.g., Mowrey et al (1991) \cite{Mow}) when the critical energy per
carbon atom (i.e., $E_{crit}/n$, where $n \geq 60$) exceed 2.7 $eV$. If
$E_{crit}/n \ge 3.5$ $eV$, then the instant fragmentation proceeds rapidly
(Mowrey et al (1991) \cite{Mow}). The average bond strengths in fullerenes
are of the order $4.5 - 5$ $eV$. From here one finds that the instant
fragmentation of the $C_{60}$ molecule can start, if the critical energy
$E_{crit} \ge 2.7 \times 60 \approx 162$ $eV$. In the present case, the
maximal excitation energy $E$ is $\le 77$ $eV$. Therefore, the effective
excitation energy per each carbon atom ($\approx$ 1.3 $eV$) is approximately
twice smaller than 2.7 $eV$. In fact, for the $C_{60}$ molecule the bulk of
the incident excitation energy is distributed among the 60 atoms in the
$C_{60}$ molecule. Moreover, if the $C_{60}$ molecules are associated (or
implanted) in some larger molecular clusters, e.g., carbon nanotubes (see,
e.g., Monthioux (2002) \cite{Month}), then the incident molecular excitation
(e.g., 77 $eV$ and 100 $eV$ excitations) can be transferred almost instantly
to/from some distant molecules. In particular, below we shall assume the
incident state of the fullerene $C_{60}$ molecule in an excited state, while
the central ${}^{235}U$ atom in its ground state. The same consideration can
be applied to any fullerene molecule which contain the central uranium atom
${}^{235m}U@C_n$ where $n$ = 84, 100, 128, etc.

In addition to the regular fullerene molecules, one can also consider the
similar ${}^{235m}U@C_{60-m}X_m$ molecular structures, where $X$ are the
non-carbon atoms (e.g., boron, nitrogen or hydrogen atoms) and $m \leq 12$
(Guo et al (1991), Hummelen et al (1995), Hultman et al (2001) \cite{Guo},
\cite{Humm}, \cite{Hult}). In general, the $C_{60-m}X_m$ molecules are
slightly less stable than the pure fullerene molecules $C_{60}$. The
molecular bond strengths usually decrease by $\approx$ 0.30 $eV$ per
substituted atom \cite{Guo}. In the case of nitrogen such a deviation can
be even $\approx$ 0.60 $eV$ per each additional nitrogen atom (Hummelen et
al (1995), Hultman et al (2001) \cite{Humm}, \cite{Hult}). However, in any
case the considered $C_{60-m}X_m$ molecules are stable and can be used for
our present purposes. Moreover, currently, the ${}^{235m}U@C_{60-m}X_m$
molecular structures are the most promising systems for the future
experiments to study the internal molecular conversion of low-energy nuclear
transition. Indeed, by varying the number of substituted atoms in the
${}^{235m}U@C_{60-m}X_m$ molecules one can change the occupation numbers
$N$ and molecular matrix element $w^{(M)}_e(n_1, \ell_1, j_1, A;\epsilon_2,
\ell_2, j_2; F)$ in Eq.(6). In some ${}^{235m}U@C_{60-m}X_m$ molecules, the
resonance conditions between the incident and final states can be obeyed
almost exactly. In contrast with this, there is no sense to apply the higher
fullerenes ${}^{235m}U@C_n$, where $n \ge 128$, in the experiments related
to the internal molecular conversion of low-energy nuclear transition. This
follows from the fact that the effective uranium-carbon distance in
fullerenes decreases with $n$. The molecular matrix elements $w^{(M)}_e(n_1,
\ell_1, j_1, A; \epsilon_2, \ell_2, j_2, F)$ for large $n$ ($n \geq 128$)
almost coincides with the corresponding atomic (${}^{235}U$) matrix element
Eq.(5). The presence of distant carbon atoms in the ${}^{235m}U@C_n$
molecule does not play any noticeable role for $n \ge 200$.

In this study our analysis was restricted to the ${}^{235m}U@C_{60}$
molecule only. Moreover, it was assumed that the outer electrons in the
incident ${}^{235}U$ atom form the $(6s_{\frac12})^2 (6p_{\frac12})^2
(6p_{\frac32})^4 (5f_{\frac52})^3 (6d_{\frac32})^1 (7s_{\frac12})^2$
electron configuration. Note that the $(5f)^3 (6d)^1 (7s)^2$ outer electron
configuration (term ${}^5L_{J=6}$, odd parity) is usually considered
(see, e.g., Avery (2003) \cite{Aver}) as the valency configuration (i.e. the
ground state) of the uranium atom. The first excited state (term
${}^5K_{J=5}$, odd parity) of the uranium atom has relatively small
excitation energy $\Delta \approx 0.077$ $eV$. In general, the valency
configuration changes drastically when uranium atom is bounded into
different molecules or implanted in various metallic alloys. Nevertheless,
it was assumed in earlier works (M\'{e}vergnies (1972), Grechukhin and
Soldatov (1976) \cite{Nev2}, \cite{GS1}) that the $(6s_{\frac12})^2
(6p_{\frac12})^2 (6p_{\frac32})^4$ configuration of deep-lying 6-shell
electrons does not change when uranium atoms form molecules (or implanted
in metals). In this study we also considered a number of cases when only the
valency configuration was varied. Briefly, our results for such cases can be
described as follows. By varying the population of the $6d-$orbitals one can
change (decrease) the decay rate constant $\lambda({}^{235m}U)$ only by 3
\%. Analogous variation for the $7s$ electrons produces significantly
smaller effect. For the $5f$ electrons the direct contribution to the decay
rate constant is less than 1 \%. On the other hand, the $5f$ electrons
penetrate some inner-lying electron shells of the uranium atom. Their
interaction with the $(6p_{\frac12})^2$ and $(6p_{\frac32})^4$ electron
shells can change, in principle, the occupation numbers of these two $6p$
orbitals. In general, the decay rate constant $\lambda({}^{235m}U)$
increases drastically, when the population of $6p$ electron orbitals
changes (decreases).

The idea to use the $5f$ electron shells in order to change the occupation
numbers of the $(6p_{\frac12})^2$ and $(6p_{\frac32})^4$ electron shells can
be extremely productive for the ${}^{235m}U@C_{60}$ molecule and other
similar molecules. Indeed, the excitation of one electron from the $6p$
shell to $5f$ shell decreases the constant $\lambda({}^{235m}U)$ by
$\approx 21$ \% in the case of the $(6p_{\frac12})$ shell and by $\approx
9.5$ \% for the $(6p_{\frac32})$ shell. The half-life of the ${}^{235m}U$
isomer increases correspondingly. In the case of two-electron excitation
from the same shells the half-life of ${}^{235m}U$ isomer increase up to
$\approx$ 40 minutes and $\approx$ 31 minutes, respectively. In any case,
this effect can be observed and measured experimentally. Note that the
excitation transitions from the $6p-$shell to the $6d-$shells of the $U$
atom are also possible, but they are less likely.

It is interesting to consider an opposite process, i.e. the nuclear
excitation of the ${}^{235}U$ nucleus by using molecular excitations in
various large molecules. Analysis of the energy level structure in the
uranium atom shows that this process can proceed only with the use of
$(5d_{\frac52})$ and $(5d_{\frac32})$ electrons in the uranium atom. The
corresponding (single-electron) energies of these electron shells are
$\approx$ 109.9857 $eV$ and $\approx$ 118.41311 $eV$, respectively. Such an
excitation can still be accumulated in the $C_{60-n}X_n$ molecule without
producing its instant destruction. The energies of the $(5p_{\frac32})$ and
$(5p_{\frac12})$ electron shells in the uranium atom are significantly
higher ($\approx$ 220.2218 $eV$ and $\approx$ 275.5916 $eV$, respectively).
Therefore, the $5p-, 5s-$ and other internal electron shells of the uranium
atom are of less interest for the considered applications. The transition of
molecular excitation to the uranium nucleus can proceed in a following way.
First, a vacancy is formed in the $5d-$electron shell of uranium atom. The
$6d-$electron is moved to the $5f-, 6d-, 7p-$, etc electron shells or to the
continuous spectra. This step requires $\approx$ 100 - 120 $eV$ of energy.
On the second step this $5d-$vacancy is filled by an electron from outer
electron shells, e.g., from the $6p_{\frac12}, 6p_{\frac32}, 5f_{\frac52},
6d_{\frac32}$ or $7s_{\frac12}$ shells. In some cases, some part ($\approx$
77 $eV$) of the energy released during this step can be used to form the
excited ${}^{235m}U$ nucleus. The rest of the energy can be either emitted
as a radiation quanta, or imparted to the ejection of an electron from one of
the outer electron shells without any photons being produced in the process
(Auger effect combined with the nuclear excitation). In particular, the case
when such a vacancy has formed in the $6p_{\frac32}$ electron shell is of
specific interest. Currently, we evaluate the total probability of nuclear
excitation (i.e. the probability of inverse internal conversion) as
$\approx$ 1 \%. This means that only 1 of 100 vacancies formed in the
$5d-$shells of the uranium-235 atoms will produce the corresponding nuclear
excitation. Unfortunately, in this study we cannot present more detailed
description of this process. Moreover, our current understanding of the
atom-molecular interaction in the ${}^{235m}U@C_{60-n}X_n$ molecule and
related compounds is far from complete. In the future, we are planning to
apply more accurate methods developed recently for relativistic calculations
of some complex molecules (Bagus et al (2000), Graaf et al (1998)
\cite{Bagus}, \cite{Graaf}). The consideration of atomic part of the problem
(i.e. the description of ${}^{235}U$ atom) must be also improved.
Nevertheless, after some improvements we hope to present a more accurate
picture of the internal conversion of nuclear transition in the
${}^{235m}U@C_{60-n}X_n$ molecules.

In conclusion, let us discuss some applications of the considered phenomena.
First, note that such a low-lying excited state ($\approx$ 77 $eV$) can be
found only in the ${}^{235}U$ nucleus. Analogous excited states in nuclei of
other uranium isotopes have significantly larger energies. Therefore, the
existence of the considered low-lying excited state in the ${}^{235}U$
nucleus can be used to separate the ${}^{235}U$ isotope from mixtures
containing various uranium isotopes, e.g., the ${}^{233}U, {}^{234}U,
{}^{236}U$ and ${}^{238}U$ isotopes. In fact, nowadays the separation of
uranium isotopes is not an actual problem. Note, however, that the internal
conversion of low-energy nuclear $\gamma-$quanta can be observed in some
other fissionable elements. If the nuclear isomers with relatively small
transition energies ($E \leq 150$ $eV$) do exist in the fissionable
${}^{247}Cm$ and ${}^{251}Cf$ nuclei, then it can be used to separate these
two isotopes. In turn, the industrial separation of these two isotopes will
be extremely beneficial for the future development of nuclear industry and
weaponry.

The second and very interesting application is related to a possibility
to control the neutron criticality (Weinberg and Wigner (1958) \cite{Wei})
of fissionable materials by changing the population of two nuclear states
(ground and considered low-lying excited states) in the ${}^{235}U$ nuclei.
To discuss this effect we shall use the method which is based on the
time-dependent diffusion model in fissionable materials (Weinberg and Wigner
(1958) \cite{Wei}). In this model the neutron propagation is given by the
simple diffusion equation, which contains corrections for: (1) the
absorption of neutrons by the nuclei in the medium, and (2) the production
of neutrons by the fissions of fissionable nuclei. Both of these quantities
are linear in the neutron flux $\Phi (\Phi = n \cdot v$, where $n$ is the
neutron density and $v$ is the mean neutron velocity). Therefore, their sum
can be written as one term and one-velocity diffusion equation takes the
following form
\begin{eqnarray}
 \frac{1}{v} \frac{\partial \Phi}{\partial t} =
 {\bf \nabla} (a^2 {\bf \nabla} \Phi) + \beta \Phi \; \; \; ,
\end{eqnarray}
where the parameter $a^2$ (the so-called "diffusion coefficient") is
\begin{eqnarray}
 a^2 = \frac{A}{\rho N_A (\sigma_f + \sigma_c + \sigma_p)
         (1 - \overline{\cos \psi})} \approx \frac{A}{\rho N_A \sigma_t
         (1 - \frac{2}{3 A})} \; \; \; ,
\end{eqnarray}
where $\sigma_f$ is the microscopic fission cross-section for the
considered element and $\sigma_c$ is the microscopic neutron absorption
cross-section. In fact, $\sigma_c$ means the so-called non-productive
neutron capture cross-section. $\sigma_p$ is the macroscopic scattering
cross-section, and $\sigma_t$ is the total neutron cross-section. Also,
in this equation $\rho$ is the macroscopic density and $N_A = 6.0221367
\cdot 10^{23}$ is the Avogadro number, while $\overline{\cos \psi} \approx
\frac{2}{3 A}$ is the so-called average cosine of neutron scattering
(Weinberg and Wigner (1958) \cite{Wei}). The parameter $\beta$ in Eq.(9)
takes the following, well known (Weinberg and Wigner (1958) \cite{Wei}) form
\begin{eqnarray}
 \beta = \frac{\rho N_A}{A} \Bigl((\nu - 1) \sigma_f - \sigma_c \Bigr)
 \; \; \; ,
\end{eqnarray}
where the parameter $\nu = \nu(E)$ is the number of neutrons released per
one fission, which is produced by a neutron with the energy $E$. The energy
dependence of $\nu(E)$ is approximated by the following linear expression
(Henkel (1964) \cite{Hen}) $\nu (E) = \nu_0 + \alpha E$, where $\nu_0$ is
the number neutrons released per fission with the thermal neutrons, and $E$
is the energy of the initial neutron (in $MeV$). For the uranium-235 we have
$\nu_0 \approx$ 2.432 and $\alpha \approx 0.100$ (Hoffman and Hoffman (1974)
\cite{Hof}).

In the case when $\beta > 0$ (i.e. $\nu \sigma_f > \sigma_f + \sigma_c$,
or $\eta = \frac{\nu \sigma_f}{\sigma_f + \sigma_c} > 1$) the intensity of
chain reaction will increase. This corresponds to the supercritical
fissionable system. For binary mixture of the ${}^{235}U$ and ${}^{235m}U$
nuclei we can write $\sigma_f = x \sigma^{(m)}_f + (1 - x) \sigma_f$ and
$\sigma_c = x \sigma^{(m)}_c + (1 - x) \sigma_c$, where $x$ is the isomer
concentration, $\sigma^{(m)}_f$ and $\sigma^{(m)}_c$ are the corresponding
neutron cross-sections of the ${}^{235m}U$ nucleus, while $\sigma_f$ and
$\sigma_c$ are the neutron cross-sections of the ${}^{235}U$ nucleus in its
ground $(\frac72)^-$ state. If there is a way to control the ${}^{235m}U$
isomer concentration $x(t)$, then it can be used to transform the
fissionable mixture to a supercritical state (or vice versa to an
undercritical state). There are a number of other applications which are
based on internal convergence of low-energy nuclear transition in the
${}^{235m}U$ nuclei. The reversibility of such a conversion in the
fullerene-based molecular structures which contain the ${}^{235}U$ atoms
allows us to consider a significantly larger number of applications. For
instance, by using low-energy molecular excitations one can produce, in
principle, the nuclear pumping in the ${}^{235}U$ sample. This means that
the nuclear properties of such a sample can be changed in the result of
molecular excitations. In conclusion, it should be mentioned that the
considered molecular conversion of nuclear low-energy nuclear transition in
the fullerene-based ${}^{235m}U@C_{n}$ molecules and related compounds
${}^{235m}U@C_{60-n}X_n$ warrants further theoretical and experimental
study. In fact, some other large molecules which contain ${}^{235}U$
atoms can also be considered. Hopefully, this work will stimulate further
experimental activity in studying of this very interesting phenomenon. If
someone is interested in performing experiments described above, please, let
me know by e-mail.

\begin{center}
  {\bf Acknowledgment}
\end{center}

It is a pleasure to thank Professor Ria Broer (Groningen, The Netherlands)
for permission to use the MOLFDIR computational package.

\end{document}